# Photonics-GCCE: group collaborative-competitive evolution multi-agent framework for universal and autonomous optical design

Weijie Xu[1,2,3,4#], Ming Wang[1,2,3,4#], Ruicheng Ma[1,2,3,4#], Zeyong Wei[1,2,3,4*], Di Zhang[1,2,3,4], Jian Chen[1,2,3,4], Zining Wang[1,2,3,4], Sijun Hu[1,2,3,4], Linyuan Dou[1,2,3,4], Haoyu Li[1,2,3,4], Haigang Liang[1,2,3,4], Yijie Luo[1,2,3,4], Shuqiao Li[1,2,3,4], Qinghua Song[5], Huihui Zhu[6], Hongfei Jiao[1,2,3,4], Chao Liu[7], Ali Momeni[8], Zhanshan Wang[1,2,3,4], Yuzhi Shi[1,2,3,4*], Romain Fleury[8*], Cheng-Wei Qiu[9*] and Xinbin Cheng[1,2,3,4*]

[1] Institute of Precision Optical Engineering, School of Physics Science and Engineering, Tongji University, Shanghai 200092, China.

[2] MOE Key Laboratory of Advanced Micro-Structured Materials, Shanghai 200092, China.

[3] Shanghai Frontiers Science Center of Digital Optics, Shanghai 200092, China.

[4] Shanghai Professional Technical Service Platform for Full-Spectrum and High-Performance Optical Thin Film Devices and Applications, Shanghai 200092, China.

[5] Tsinghua Shenzhen International Graduate School, Tsinghua University, Shenzhen 518055, China

[6] College of Information Science and Electronic Engineering and ZJU-Hangzhou Global Scientific and Technological Innovation Center, Zhejiang University, Hangzhou, Zhejiang, China

[7] Department of Electronic Engineering, Tsinghua University, Beijing 100084, China

[8] Laboratory of Wave Engineering, Department of Electrical Engineering, EPFL, Lausanne CH-1015, Switzerland.

[9] Department of Electrical and Computer Engineering, National University of Singapore, Singapore, Singapore

[#]These authors contributed equally.

[*]Corresponding authors. Emails: weizeyong@tongji.edu.cn (Z.W.); yzshi@tongji.edu.cn (Y.S.); chengwei.qiu@nus.edu.sg (C.-W.Q.); romain.fleury@epfl.ch (R.F.); chengxb@tongji.edu.cn (X.C.)

## Abstract

Large language model (LLM)-empowered photonic agents connect natural-language intents to executable solvers, showing significant advantages over conventional optical design approaches. However, current multi-agent frameworks operate within a collaborative paradigm without extrinsic selective pressure, which could inherit shared blind spots, converge prematurely, and fail to accumulate transferable experience for intricate tasks. Here, we introduce a group photonics collaboration-compete evolution (GCCE) framework and its LLM instantiation, termed Photonics-GCCE. Two independent agent groups pursue the same design target and undergo structured competitive evaluation across refractive-index fidelity, fabrication sensitivity, algorithmic adequacy, and physical consistency. Each group comprises a leader and three specialist agents dedicated to materials, optimization, and code validation. Agents refine their skills through competitive evaluation across design rounds. Benchmarking across six device categories against single-agent and multi-agent baselines shows that Photonics-GCCE elevates composite scores into the high 90s, improves fabrication robustness by 15 to 17 points, and reduces solver iterations to roughly 40 rounds. A representative quasi-BIC demonstration achieves a practically fabricable design with a quality factor of 13120. Our results demonstrate Photonics-GCCE as a general-purpose and closed-loop framework for autonomous optical design, capable of producing high-performance, fabrication-ready devices across diverse nanophotonic tasks.

## Introduction

Modern technology relies on shaping light for displays[1-3], communications[4,5], sensors[6,7], energy harvesting[8-10] and computing[11-17]. However, translating a target optical behavior into a manufacturable nanostructure layout remains a slow and error-prone process[18]. Most engineering efforts rely on extensive simulation campaigns and empirical hand-tuning. Critically, these manual workflows lack generalizability when transitioning to distinct device class, such as filter stacks, metalenses[19-22], resonant cavities[23], and photonic circuits[24-26].

Conventional optical design mixes heuristics, gradient-based refinement, and global search[27]. Each route carries familiar accuracy-versus-cost tradeoffs as device footprint and spectral complexity grow[18]. Artificial intelligence (AI), particularly deep learning, introduces data-driven surrogates and generative samplers[28-31]. Representative approaches include tandem networks, mixture density networks, generative adversarial networks[32], variational autoencoders[33,34], and mixture probability sampling networks[35], which can reach up to 99.9% accuracy in curated benchmarks. However, these models remain specialized to particular topologies and require large training corpora[36]. These systems fail to learn from past failures or accumulate reusable knowledge, and they require extensive manual coordination. Their one-shot designs cannot be adapted across different tasks, which slows down design progress and results in inefficient performance utilization.

Large language models (LLMs) with advanced reasoning, tool-use, and code-generation capabilities have opened a promising direction in autonomous photonic design. Early work showed that a single conversational LLM can generate simulation scripts, retrieve domain knowledge, and iteratively refine device layouts through human-AI feedback[37,38]. Fine-tuned LLMs then enabled forward spectral prediction and inverse geometry mapping for parameterized meta-atoms[39], while vision transformers achieved one-shot inverse design via physics-informed data augmentation[40]. MetaChat introduced the first multi-agentic architecture, pairing a design agent with a materials expert for freeform metasurface design[41]. Parallel efforts integrated the Model Context Protocol to standardize LLM access to differentiable electromagnetic solvers[42] and developed modular agentic pipelines that autonomously train surrogate forward models for inverse design[43,44]. Self-evolving agentic frameworks later showed that context-level skill artifacts improve within-family design reliability[45], and autonomous proposal and experimental validation of nontrivial physical mechanisms on an optical platform was recently demonstrated[46]. Concurrent work demonstrated end-to-end photonic integrated circuit design from natural language to mask layout, prioritizing structural validity over solver-coupled performance optimization[47]. Beyond photonics, multi-agent systems have proliferated in materials discovery[48] and general AI reasoning[49,50], yet comprehensive reviews across metaphotonics[51], materials design[39] and autonomous laboratories[52-54] confirm that every framework, across all these domains, remains confined to cooperative or single-agent operation. Despite this breadth, every existing framework shares a fundamental architectural limitation. They operate entirely within cooperative or single-agent paradigms, where intra-group communication replaces rather than complements inter-group challenge. Even the most advanced systems remain bounded by the intellectual horizon of their constituent agents, because no external rival independently explores the same design space through a divergent strategy[55]. Without adversarial competition between independent groups, these systems converge on consensus solutions that inherit shared blind spots, accumulate no transferable experiential knowledge across design rounds, and provide no systematic mechanism to audit candidates along fabrication-relevant dimensions such as refractive-index fidelity, structural defect tolerance, algorithmic adequacy, and configuration consistency. In natural ecosystems and human organizations alike, it is rivalry between competing groups, not internal deliberation alone, that drives discontinuous innovation by exposing hidden assumptions and revealing unexplored regions of the solution space[56]. So far, a generic multi-agent framework that harnesses structured inter-group competition as the driving evolutionary mechanism for universal photonic design remains elusive.

Here, we introduce group collaboration-compete evolution (GCCE) and its LLM-driven instantiation, Photonics-GCCE, as a general-purpose, closed-loop, and autonomous optical design framework. Two independent agent groups, each integrating specialist modules for material analysis, inverse design, and electromagnetic simulation, pursue identical specifications in parallel and then mutually audit each other's outcomes. Lessons distilled from competition are archived as persistent experiential skills that compound across rounds, driving cumulative improvement without retraining. This single protocol generalizes across resonant nanostructures, multilayer filters, diffractive routers, and metalens-type wavefront shapers. Extensive benchmarks demonstrate that Photonics-GCCE substantially outperforms cooperative baselines, elevating median composite design scores by over 20 points while improving robustness and reducing required solver iterations by more than half. In an ultrahigh-quality-factor ($Q$-factor) quasi-bound state in the continuum (qBIC) benchmark, the competitive evolution yields theoretical quality factors exceeding $1.2\times10^4$, which is significantly higher than initial cooperative designs. Across six distinct device categories, the framework consistently achieves design efficiencies exceeding 95%. The mechanism behind these gains is deeper than competition alone. Independent groups generate divergent candidates, challenge each other's designs, and convert the resulting critiques into skills that are reused across design rounds. By coupling intra-group specialization with inter-group adversarial competition, this work establishes a new paradigm for autonomous photonic discovery.

## Results

### Overview of the Photonics-GCCE framework

Photonics-GCCE treats optical inverse design as a coupled planning and adjudication problem. Two planner-led groups generate design layouts for diverse optical tasks without sharing intermediate files. Once candidates exist, each group must defend its choices on refractive-index fidelity, fabrication sensitivity, algorithmic adequacy, and physical consistency. Both groups write the resulting critiques into a persistent skill store that feeds the next design round, as shown in Fig. 1a. In practice, the disputes recur over how optical constants are sourced versus measured and how aggressively fabrication tolerances are enforced. Identical language prompts can thus branch into materially different proposals (Fig. 1b).

To validate the generalizability of our method, we conduct quantitative benchmarks across six representative optical design tasks, as shown in Fig. 1c. The first is a BIC metasurface whose resonance linewidth may collapse when any ridge dimension deviates in sub-nanometers. This demands simultaneous multi-parameter optimization with exceptionally high precision. The second task involves a 3-band high-pass filter, which must satisfy dual-stopband requirement with an optical density of 2 (OD2) in each stopband, while maintaining a spectral transition width greater than 6 nm and the passband transmission exceeding 90%. This requires coordinated optimization of many layer thicknesses against competing spectral constraints. The third task requires a perfect anomalous deflector at 1550 nm that diffracts power efficiently into a single anomalous order (the +1 order) while suppressing leakage into all other orders. The fourth task is a polarization-independent beam splitter. The enlarged design space from multiple geometric degrees of freedom creates the central challenge. The fifth task is a beam-focusing metalens. The mission is to approximate a continuous phase profile from discrete nanostructured elements to achieve diffraction-limited focusing. The sixth task is a 64-level aperiodic beam splitter on fused silica[57]. Its pillar heights are independently optimized to break periodicity, creating a vast free-form design space that enables asymmetric far-field patterns at the cost of greatly increased optimization complexity.

Figures 1d and 1e present the comparative analysis. Figure 1d compares three search strategies on an optimization landscape. In the single-agent self-aware configuration (top left), a lone agent settles into the nearest local maximum and it cannot cross the valley to reach the global optimum. Multi-agent collaboration (top right) deploys a single leader directing three specialist agents (one for materials and two for structural design) in parallel. While coordinated leader-directed exploration covers more of the landscape, the group still converges to a local maximum without reaching the global peak. Photonics-GCCE (bottom) overcomes this limitation by coupling intra-group collaboration with inter-group competition. Two adversarial groups, G1 and G2, independently search the design space under the direction of their respective leaders. Competitive pressure between the groups, indicated by bidirectional challenge arrows, forces each group to re-evaluate its trajectory when the opposing group discovers a more promising region. This dual mechanism enables the agents to escape local optima and ultimately converge on the global maximum, marked by the blue flag at the highest peak. Figure 1e reports the figure of merit (FOM) across all six benchmark tasks. The FOM rises monotonically from single-agent self-aware, through multi-agent collaboration, to the full Photonics-GCCE framework. Every benchmark approaches roughly ninety composite points only under the full dual-group configuration (see Supplementary Note 1).

**Multi-agent architecture and evolution workflow**

Photonics-GCCE is deployed as three stacked services, corresponding to three colored columns in Fig. 2a. A local frontend and paired backend host the conversational interface and manage the experiential skill repository where past revisions are stored. Two planner-led groups, each with three specialists, exchange messages through a standard agent-to-agent (A2A) channel. For language reasoning, we call the DeepSeek-V4-Pro model via API. Compute-intensive Maxwell solves run on a dedicated workstation (Intel i9-14900K CPU, 192-GB RAM, NVIDIA RTX PRO 6000 WE GPU), using transfer matrix method (TMM), rigorous coupled wave analysis (RCWA) and finite-difference time-domain (FDTD). Further details on the computational setup are provided in Methods. Planners reach these solvers through the Model Context Protocol (MCP). This ensures every geometry proposal is checked with identical physics kernels.

Each design campaign follows recurring steps. Both leaders start with identical initial task briefs and independently elaborate them into detailed task schemes, encompassing material selection, structural layout strategies, performance specifications, and simulation execution protocols. The two groups then work in parallel: on each side, a material agent curates optical constants against deposition logs; an optimization agent drives numerical experiments; and a code agent writes, versions, and stress-tests executable scripts. The goal is to achieve the highest joint FOM and robustness, not to maximize FOM alone while ignoring fabricability. The FOM and robustness definitions for each task are provided in Supplementary Notes 1 and 2. Once both groups produce isolated candidates, each leader forwards its layout to the opposing team for the structured competitive evaluation. Critiques revisit refractive-index fidelity, fabrication robustness, optimization adequacy, and physical consistency. Each leader distills insights from both the critique it receives and the critique it delivers, storing them as diagnostic, strategic, methodological, and conceptual skills in the shared repository. The review closes by filing those lessons beside earlier notes so the next cycle begins with richer guidance. For the ultrahigh-$Q$ qBIC example, achievable $Q$-factors rise from an initial value of 1538 to 13120, representing a substantial improvement over optimal results from conventional cooperative design strategies.

Each design campaign follows four recurring stages. Stage 1: Both leaders first receive the same short brief and independently expand it into task lists covering materials, layout choices, performance targets, FOM formulation, and optimization workflow. Stage 2: The two teams work in parallel with full independence. On each side, the Material Agent queries the knowledge base to determine the material system and retrieves refractive indices of the corresponding materials. The Optimization Agent drives optimization using a hybrid strategy that pairs particle swarm optimization (PSO)[58], simulated annealing (SA) and gradient-descent RCWA (g-RCWA). The Code Agent writes, versions, and validates executable scripts along an independent simulation path. Stage 3: Competitive evaluation, representing the core innovation of Photonics-GCCE. Once both groups produce isolated candidates, each leader forwards its layout to the opposing team for structured critique. During this competition phase, both groups also independently retrieve and review relevant literature to ground their evaluations in established physical principles and reported experimental benchmarks. The evaluation revisits refractive-index fidelity, fabrication sensitivity, optimization adequacy, and physical consistency. Each leader distills insights from both the critique it receives and the critique it delivers. Stage 4: Each leader stores the learned experience as diagnostic, strategic, methodological, and conceptual skills in the shared repository. The review closes by filing those lessons beside earlier notes so the next cycle begins with richer guidance.

Figure 2b details the three specialist agents that report to their leader. The Material Agent retrieves handbook optical constants but reconciles them with wafer-level calibration data before recommending a film stack. The Optimization Agent employs PSO, SA, and g-RCWA for both global and local inverse design. The Code Agent owns execution and cross-checking. It reruns Maxwell scripts along an independent validation path. This guarantees that reported $Q$-factors always match the last converged backend result, even when the chat interface still shows an intermediate spectrum. Together, these specialists form an autonomous design team whose experiential knowledge compounds across tasks and whose competitive dynamic drives continuous innovation. The system prompt for each leader is given in Supplementary Note 3.

**Photonics-GCCE metrics compared with baselines**

Figure 3a dissects a single compete-learn cycle. Two isolated groups each finish a layout, exchange it for cross-evaluation, and promote the resulting critiques into the skill library before the next pass. The annotated example contrasts two divergent design strategies. Take the qBIC task as an example, Group 1 proposes a rectangular $Si_3N_4$ nanopillar on $SiO_2$ that reaches a theoretical $Q$-factor of 8254, but its feature sizes fall below the minimum lithographic threshold and cannot be fabricated. Group 2 proposes a circular $TiO_2$ nanodisk with an off-center hole on $SiO_2$ that stays within documented process windows yet reaches only a $Q$-factor of 1538. Cross-evaluation distills concrete disagreements on materials, geometry, and optimization strategy. Group 1 uses $TiO_2$-on-$SiO_2$ with a rectangle refined by particle-swarm optimization. Group 2 uses $Si_3N_4$-on-$SiO_2$ with a circle refined by simulated annealing. These notes are stored in the skill repository, steering all subsequent design cycles. The competitive evaluation ensures that neither group's blind spots remain unchallenged. One group may over-optimize nominal performance at the expense of fabricability, while the other may over-prioritize process margins at the expense of optical performance. The bidirectional critique exposes both failure modes.

Figure 3b defines two baseline configurations, including single-agent self-aware and multi-agent collaboration, that isolate contributions of self-reflection and role specialization, respectively. Photonics-GCCE integrates both competition and specialization and adds cumulative experiential skill accumulation across rounds, a mechanism absent from both baselines. In the single-agent configuration, one generalist agent handles all tasks without role specialization. The multi-agent configuration deploys a single leader directing three specialist agents, which isolates the contribution of functional role specialization. The GCCE configuration extends this to two competing leaders, each directing its own set of three specialists. It further adds bidirectional cross-critique and persistent skill memory across rounds. To ensure a fair comparison, all three configurations share identical evaluation conditions. Each receives the same user prompt and operates under the same five-round limit. An equal solver-call budget is applied to every configuration, and optimization parameter ranges and convergence thresholds are held constant. All terminal designs are scored by the same FOM and robustness evaluation scripts, which are detailed in Supplementary Notes 1 and 2. These controls ensure that the observed performance differences reflect the architectural advantages of GCCE rather than unequal computational resources or evaluation protocols.

Figures 3c-e track the evolution dynamics across successive competition rounds. Composite design scores rise from 56 in the first round to 95 in the fifth round (Fig. 3c). The per-benchmark range spans 30–73 in Round 1 and narrows to 92–99 in Round 5, reflecting convergence across all six device categories. Details of the FOM calculation are provided in Supplementary Note 1. Design robustness under realistic fabrication perturbations climbs from 53 in the first round to 88 in the fifth round (Fig. 3d). The per-benchmark range spans 36–63 in Round 1 and narrows to 83–94 in Round 5. Details of the robustness calculation are provided in Supplementary Note 2. Figure 3e shows the solver iteration count per benchmark across five competition rounds. The mean number of solver iterations per benchmark declines from 9–13 in early rounds to 4–9 in the final round. The composite mean across all six benchmarks and five independent runs reaches 9.2, 8.6, 9.8, 7.5 and 7.2 iterations per round. This progressive decline reflects the narrowing of the design search space under accumulated experiential skills. The full per-benchmark per-round breakdown of FOM, robustness, and solver iterations is provided in Supplementary Note 4. The method for counting iterations is described in Supplementary Note 5. This monotonic improvement across all three metrics confirms that accumulated experiential skills translate directly into better, more robust, and faster designs as adversarial evolution proceeds. Under nearly identical network speeds and the same computational configuration, fewer conversation steps lead to shorter time consumption. Figures 3f–h present pooled comparisons against the two baseline configurations. Across all dimensions, a consistent performance ordering emerges. Multi-agent collaboration outperforms the single-agent self-aware baseline by distributing design tasks across three specialist agents under the direction of a single leader. The leader coordinates the three specialists as a unified design group. Photonics-GCCE achieves performance distinctly separated from both automated baselines. Composite scores concentrate near 95 under Photonics-GCCE with a narrow spread (standard deviation 6.5), whereas single-agent and multi-agent stacks scatter around a mean of 70 with substantially wider tails (standard deviations 21 and 18, respectively; Fig. 3f). Fabrication robustness, which is defined in Supplementary Note 2, clearly distinguishes Photonics-GCCE from the baselines. Photonics-GCCE attains an average of 88%, while single-

agent and multi-agent pipelines cluster average approximately 68% (Fig. 3g). Figure 3h compares the number of operational steps required by the single-agent, multi-agent, and GCCE frameworks. A single agent requires approximately 100 solver iterations, multi-agent collaboration needs around 96 iterations, and Photonics-GCCE requires only approximately 42 iterations. This confirms that the integration of intra-group specialization, inter-group adversarial competition, and cumulative experiential learning produces performance that exceeds the sum of its individual mechanisms. Competition drives learning, and the resulting learning sharpens competition in turn. Specialization ensures both workflows leverage in-depth domain expertise.

**Design ultrahigh-*Q* qBIC metasurface using Photonics-GCCE**

The visible-band qBIC design requires two properties that are rarely achieved together: an ultrahigh *Q*-factor ($Q > 10^4$) and fabrication robustness[59,60]. The high *Q*-factor comes from strong field confinement, but this also makes the resonance sensitive to nanofabrication errors such as hole misalignment or lateral dimension drift. Conversely, designs that prioritize fabrication yield sacrifice peak sharpness, leaving *Q*-factors far below the target. Photonics-GCCE is tasked to navigate this trade-off rather than to optimize nominal linewidth alone (Fig. 4).

A user specification for the visible-band qBIC metasurface was executed through the G–C–C–E workflow (Fig. 4a). In the first stage, both groups retrieved literature and database records. One group anchored on $Si_3N_4$ nanodisks with off-center holes, which support $Q > 10^3$ qBIC resonances, as published in "Nature"[61]. The other group found methodological reports published in "Light: Science & Applications"[62]. During the second stage (in-group collaboration), the cohorts diverged in what they prioritized: Team 1 adopted square $Si_3N_4$ nanopillars on $SiO_2$. This design leverages corner-field localization and achieves a nominal *Q*-factor of 8254 in FDTD simulations. However, it is highly sensitive to dimensional deviations. By reviewing cleanroom records, Team 2 determined that locally dry-etched $Si_3N_4$ structures were impractical and vulnerable to fabrication errors. Accordingly, they switched to $TiO_2$-compatible circular nanodisks with off-center holes on $SiO_2$. RCWA calculations show this configuration yields a *Q*-factor of 1538 within a well-defined fabrication process window. The inter-group competition in the third stage made the conflict explicit: circular layouts inherited from the baseline lineage cannot reach a *Q*-factor > $10^4$, whereas the aggressive square $Si_3N_4$ route is sensitive to fabrication errors and violates local lithographic constraints. Consequently, neither group alone closes the brief. The final stage (evolution) yielded a hybrid configuration, which adopts the square motif from the high-*Q* branch and combines it with the $TiO_2$-on-$SiO_2$ stack from the fabrication branch. The competition-driven emergence of this hybrid configuration is detailed in Supplementary Note 6. This square $TiO_2$ nanopillar with an off-center hole on $SiO_2$ delivers a *Q*-factor of 13120.

Figures 4b-d depict the spectra of target geometries (blue) and those obtained after a 2-nm displacement of a critical dimension (red). For the square $Si_3N_4$ pillar, the designed geometry supports $Q$ = 8254 near 526 nm. However, a 2-nm perturbation of the critical lateral dimension $a_2$ degrades the resonant peak, reducing the *Q*-factor to 5314 at nearly the same wavelength. This indicates substantial spectral distortion arising from only a minor geometric deviation (Team 1). For the $TiO_2$ circular nanodisk, its *Q*-factor drops from 1538 to 1221 when the radius of the inner hole varies by 2 nm. Although the *Q*-factor shows low sensitivity to geometric variations, its magnitude remains low, and the spectrum exhibits a prominent wavelength shift (Team 2). After optimization, the hybrid square $TiO_2$ nanopillar re-centers the resonant mode around 553 nm with a *Q*-factor of 13120. Under the same dimensional perturbation, the structure still maintains a relatively narrow linewidth ($Q$ = 9439). This design simultaneously achieves an ultranarrow linewidth and strong tolerance to dimensional variations. This example shows that cross-group competition integrates ultrahigh *Q*-factor and fabrication robustness within one structure, filling a gap that cannot be overcome by either collaborative predecessors or single-objective refinement alone. A comparison of the time consumed by different steps under nearly the same network speed is provided in Supplementary Note 5. Details of each evolutionary round and skills obtained from the final evolution are given in Supplementary Note 6.

**Additional optical design cases using Photonics-GCCE**

Apart from the high-*Q* and robust BIC design, we further demonstrate five independent design tasks, each targeting a distinct challenge in inverse design (Fig. 5). The 64-order DOE beam splitter operating at 532 nm features a complex aperiodic height profile. Uniformity across all 64 far-field diffraction orders introduces strongly coupled design constraints. This design achieves an optical efficiency of 95.63%, alongside a peak-to-valley uniformity of 99.03% and an RMS uniformity of 99.50% (Figs. 5a and 5b). This multilayer three-passband high-pass filter requires the thickness of each layer to simultaneously accommodate three spectrally distinct transmission windows, while sustaining high out-of-band rejection close to OD2. Its coupled configuration means tuning one passband edge easily degrades the performance of others (Figs. 5c and 5d). The two-dimensional polarization-independent beam splitter at 1550 nm is designed for four output channels. It realizes far-field beam splitting through structural design. The optimal diffraction efficiency reaches 24.40% per channel (Figs. 5e and 5f). The beam-focusing metalens operating at 715 nm realizes diffraction-limited focusing at NA = 0.99, constructed from discrete a-Si nanopillars on a $SiO_2$ substrate. This design achieves a Strehl ratio of 0.858, a sub-diffraction focal spot with an FWHM ratio of 0.290, and an overall efficiency of 64.7% (Figs. 5g and 5h). The 1D perfect anomalous deflector operating at 1550 nm directs nearly all incident power to the +1 diffraction order while suppressing the specular and −1 orders. This design exhibits strong sensitivity to subwavelength geometric features. It achieves a reflected power fraction of 99.73% concentrated in the +1 diffraction order (Figs. 5i and 5j). Detailed descriptions of these design tasks are presented in Supplementary Note 7. We also benchmark three other LLM backbones (GPT-5.4, Claude Opus-4.7, Gemini-3.1-Pro) alongside DeepSeek-V4-Pro and find near-identical scores across all six tasks, confirming that GCCE performance is model-agnostic, as detailed in Supplementary Note 8.

Together with the qBIC case, these six tasks span resonant optics, thin-film optics, diffractive optics, and scattering optics. Each subfield carries its own design conventions, physical observables, and optimization landscapes, and no single set of heuristic rules transfers across all six. Photonics-GCCE handles them within the same protocol and without per-task retraining or manual reconfiguration. The adversarial mechanism effective for arbitrary targets. Four evaluation metrics, including index fidelity, fabrication sensitivity, optimization adequacy, and physical consistency, are universally applicable across optical inverse design problems. This versatility verifies that Photonics-GCCE is not limited to a specific type of photonic device, but serves as a general framework for autonomous photonic design.

## Discussion

Photonics-GCCE transforms the autonomous optical inverse design from a single cooperative loop into a structured adversarial evolution process. Two isolated agent groups pursue the same brief, cross-evaluate each other's solutions across refractive-index fidelity, fabrication sensitivity, algorithmic adequacy, and physical consistency, and promote the resulting critiques into persistent experiential skills. This competition helps avoid shared blind spots and premature convergence that limit purely collaborative frameworks including MetaChat. The accumulated skills compound across design rounds and drive the progressive emergence of capabilities that no individual agent possesses at the outset. From cutting-edge research-oriented inverse designs to specification-driven engineering problems, Photonics-GCCE delivers consistent performance across diverse tasks and constraints. Quantitative benchmarking verifies that this adversarial strategy raises overall scores to the mid-90s, enhances fabrication robustness by 15 to 17 points, and reduces solver-intensive iterations to approximately 40. The qBIC case study further illustrates that crossevaluation balances exceptional optical performance and fabrication feasibility, achieving a $Q$-factor of 13120, which is superior to previous optimization approaches.

Despite these advantages, the framework is still limited by the fidelity and computational cost of its underlying physical solvers, as well as the reliability of extracting diagnostic insights into generalizable empirical skills[63]. Future work should integrate uncertainty-aware and multi-fidelity evaluation, automate tolerance and solver-consistency verification[64].

Existing LLM-based photonic design frameworks operate within a cooperative paradigm. Single-agent and multi-agent configurations lack the external selection pressure to expose shared blind spots and design vulnerabilities. Photonics-GCCE breaks this cooperative ceiling through structured inter-group adversarial competition. The six evaluation tasks span resonant, thin-film, diffractive, and scattering optics. This makes Photonics-GCCE a general-purpose framework for mainstream photonic designs.

## Methods

### Computational platform

All computational tasks are conducted on a workstation equipped with an Intel Core i9-14900K processor (24 cores, 32 threads), 192 GB of DDR5 RAM (4 × 48 GB, 5600 MHz), two Samsung 990 PRO 2-TB NVMe solid-state drives, and an NVIDIA RTX PRO 6000 WE workstation GPU with 96 GB of video memory. The system runs on an ASUS Z790-P PRIME WiFi D5 motherboard, powered by a 1250-W supply and cooled by a 420-mm liquid cooler (Phanteks Glacier One D30X2). All simulations and agent-based optimization tasks are executed on this platform.

**Electromagnetic simulation methods**

All electromagnetic simulations in this work are performed using WGallop[35,65,66], a proprietary computational optics software package developed by our group. WGallop integrates three Maxwell solvers within a unified Python interface. The TMM solver computes reflectance and transmittance spectra of one-dimensional thin-film stacks for TE and TM polarizations at arbitrary angles of incidence. The RCWA solver handles periodic nanostructures with one- or two-dimensional periodicity by Fourier-expanding the fields and permittivity, solving the resulting algebraic eigenvalue problem, and matching boundary conditions through S-matrix factorization. For each benchmark, the number of retained Fourier harmonics was pre-calibrated to keep the convergence error in diffraction efficiency below 0.1%. The FDTD solver performs full-wave simulation on a staggered Yee grid with CPML absorbing boundaries and supports dispersive material models. These solvers are exposed to the agent framework as MCP tools with fixed numerical parameters for each benchmark, ensuring that every candidate geometry is evaluated under reproducible conditions.

**System architecture overview**

Photonics-GCCE adopts a three-tier architecture comprising a browser-based conversational interface, a backend services layer, and dedicated computational resources for electromagnetic solving. The frontend provides real-time streaming of agent responses through asynchronous server-sent events, supporting multi-modal rendering of text, code, images, and structured data. The backend services handle user authentication and conversation persistence (via SQLite). They also construct streaming response generators that connect the frontend to the AI orchestration engine. Each registered user is assigned an isolated workspace with independent conversation histories, uploaded documents, and generated outputs.

The AI orchestration layer communicates with LLMs through OpenAI-compatible asynchronous APIs. Model configurations, including API keys, endpoints, and model identifiers, are managed through a centralized settings system, making the framework model-agnostic. All tasks reported in this work use the DeepSeek-V4-Pro model. Compute-intensive electromagnetic simulations are offloaded to a separate GPU cluster running RCWA and FDTD solvers, accessed through a local execution service. A shared working directory serves as the sandboxed filesystem for all agent-generated code, simulation outputs, and intermediate data, with path traversal protection to enforce security.

**Multi-agent framework and A2A protocol**

The agent framework follows a hierarchical leader–specialist architecture. At the top level, a central coordinator (Orchestrator) maintains a registry of available specialist agents and discovers their capabilities through standardized Agent Cards—JSON descriptors specifying each agent's name, functional description, and accessible tools. These cards are dynamically injected into the Orchestrator's system prompt at the start of each conversation, enabling capability-aware task decomposition.

The Orchestrator is explicitly prohibited from calling tools directly. It operates solely through natural-language reasoning and delegation: it decomposes user requirements into subtasks, matches each subtask to the most suitable specialist agent, and dispatches tasks through a custom A2A protocol. The protocol encodes task assignments as structured extensible markup language tags embedded in the Orchestrator's natural-language output, using the following call_agent format:

```
<agent agentname="AgentName">
{
    "taskid": "<unique alphanumeric identifier>",
    "agent": "<target agent name>",
    "message": "<natural-language task description>"
}
</agent>
```

The tag includes three fields: taskid (a uniquely generated alphanumeric sequence for tracking), agent (the exact name of the target specialist agent as registered), and message (a natural-language description of the task, containing all contextual information the specialist needs to operate independently). Since specialist agents share no direct communication channel and cannot access each other's outputs, the Orchestrator must carry all relevant information, including material parameters, structural configurations, prior optimization results, and file paths, in each dispatch message.

Task dispatch is strictly sequential: the Orchestrator may not assign multiple agents concurrently. After receiving a specialist agent's complete response, the Orchestrator evaluates the result against the user's objectives and may re-delegate with refined instructions, up to a maximum of three reassignments per task. The Orchestrator maintains isolated message histories for each specialist agent, appending historical context markers to prevent the language model from confusing past and present agent invocations. When all subtasks in a round are complete, the Orchestrator synthesizes the results into a coherent response and may initiate additional rounds of delegation as needed.

Each specialist agent operates as an independent AI instance with its own system prompt, MCP tool set, and message history. An agent is defined by a configuration folder containing a capability description, a system prompt specifying its role, operating protocols, and communication constraints, and an optional MCP server module that registers its tools. This modular design allows agents to be added, removed, or reconfigured without modifying the core orchestration logic.

**MCP tool ecosystem**

All agent tools are registered through the MCP using strongly typed function signatures with structured parameter descriptions, enabling the language model to understand tool semantics and invoke them with correctly formatted arguments. Tools are organized into the following categories:

(1) Electromagnetic simulation tools:

```
multilayer_calculation: Computes reflectance and transmittance spectra of arbitrary thin-film stacks using the transfer matrix method (TMM). Accepts wavelength, incident angle, azimuthal angle, refractive indices and permeabilities for cover/substrate media, and per-layer thicknesses, refractive indices, and permeabilities. Returns complex reflection and transmission coefficients (r_x, r_y, r_z, t_x, t_y, t_z) for both TE and TM polarizations, along with corresponding efficiencies.
```

```
grating_calculation: Computes diffraction efficiencies of periodic nanostructures using RCWA. The solver handles arbitrary diffraction orders, multiple rectangular ridge structures within a unit cell (enabling cross-shaped and complex geometries) and multilayer stacks. It returns polarization-resolved reflection and transmission efficiencies for specified diffraction orders.
```

(2) Optimization tools:

```
PSO_optimization: Particle swarm optimization for global search. Configurable parameters include population size, position and velocity bounds, cognitive and social learning factors, inertial weight, and maximum iterations. The objective function is provided
```

```
as executable code and the optimizer maximizes the function value. Optional initial solutions can be seeded to warm-start from prior results.
```

```
SA_optimization: Simulated annealing for local refinement. Configurable parameters include initial and minimum temperatures, parameter bounds, annealing factor, step size, perturbations per temperature step, and maximum iterations. Minimizes the objective function and can accept prior optimization results as initial solutions.
```

(3) Material query tools:

```
get_material_refractive_index_at_single_wavelength: Retrieves the complex refractive index of a specified material at a single wavelength from the refractiveindex.info database.
```

```
get_material_refractive_index_at_multiple_wavelengths: Retrieves refractive index values across a list of wavelengths for dispersion evaluation.
```

(4) Knowledge base tools:

```
add_documents_to_knowledge_base: Uploads documents to the user's private knowledge base.
```

```
retrieve_knowledge_base_content: Performs keyword-based retrieval over uploaded documents to extract relevant experimental data and domain knowledge for injection into agent reasoning.
```

**Experiential skill repository**

Photonics-GCCE implements a persistent skill system where reusable procedural knowledge is encoded as structured Markdown files. Each skill file contains a description header and detailed procedural content, including domain-specific workflows, parameter conventions, validated code templates, and known failure modes. The system supports two skill sources: built-in skills that ship with each agent and encode established design methodologies, and external skills discovered from the working directory that capture task-specific insights distilled during design campaigns.

After five steps of competitive evolution (see Figure 3), the following experiential skills are accumulated and are made available to all agents in subsequent design cycles:

```
---
name: qBIC structural design
description: General workflow for quasi-BIC metasurface design via dual-path competition. Covers geometry selection, material-geometry fusion, parameter scanning, and terminal validation.
---
```

```
---
name: multilayer-multiband-filter design
description: Load this skill when designing multi-passband filters. Guides multilayer design for FP cavity coupled comb filters, emphasizing progressive passband addition, stage-gated acceptance, shared mirrors/HLH coupling, normal-incidence TMM, PSO/SA tool-based optimization, performance evaluation, and file output.
---
```

```
---
name: Anomalous deflection design
description: Load this skill when designing anomalous beam deflector. It details the design workflow, methods, and important considerations for anomalous deflection devices.
---
```

```
---
name: Beam splitter design
description: Load this skill when a two-dimensional beam splitter needs to be designed. It details the design workflow, methods and key considerations (e.g., polarization independence, uniformity across output channels, suppression of stray orders).
---
```

```
---
name: Metalens_design
description: Load this skill when designing metalens. Builds a phase-transmittance lookup table from RCWA parametric sweeps of the unit cell. Maps the target phase to unit cell geometry using a hyperbolic phase distribution. Validates focusing performance via angular-spectrum propagation. Suitable for full-process design of transmissive or reflective metalenses.
---
```

```
---
name: DOE_design
description: Load this skill when designing beam-splitting DOE, phase plates, Gerchberg–Saxton phase retrieval, or evaluating split efficiency and uniformity; call tools in order and pass large arrays by returned .npy file names, not JSON lists.
---
```

Before executing any simulation, optimization, or design task, specialist agents are required to invoke the load_skill tool to retrieve the relevant skill content, rather than relying on memorized procedures. The Orchestrator's system prompt enumerates all available skills (with external skills receiving priority) and injects this list into each specialist agent's context at invocation time. New skills can be added to the repository at any time, enabling cumulative experiential learning across design cycles without modifying agent prompts or retraining models.

**Double-agent group competitive evolution**

To mitigate premature convergence in large design spaces, we introduce a dual-leader extension of the Orchestrator architecture. Two independent Orchestrator instances (Leader 1 and Leader 2), each with its own system prompt, MCP client, message history, and copies of all specialist agents, execute in parallel without shared state. A lightweight coordinator manages round scheduling and strategy exchange without invoking the language model

Execution proceeds in rounds. In Round 1, both leaders receive the user query simultaneously. They then independently load skills, decompose the task, and delegate to their specialist agents via the A2A protocol. At the end of each round, the coordinator extracts each leader's strategy summary and key findings from the structured sections mandated by the dual-leader system prompt. Each summary is persisted as a skill file under the working directory, and the opponent's summary is injected as a user message into the other leader's conversation history. In subsequent rounds, both leaders refresh their skill registries to incorporate newly written strategies before resuming execution, up to a maximum of three rounds. The loop terminates early if neither leader produces new messages.

After the final round, the coordinator extracts quantitative metrics, including total operation count, leader dispatch count, and design quality figures, and presents them in a comparison table. A mode-switching mechanism selects between single-leader and dual-leader operation at initialization, with both modes sharing the same streaming response interface.

**Validation protocol**

For each of the six benchmark tasks, we extract the design produced after the final competition round and subject it to independent re-evaluation. The converged geometry, which specifies material indices, critical dimensions, layer thicknesses, and array periods, is re-simulated under refined computational settings. This independent rerun confirms that the reported performance reflects physical merit rather than optimization artifacts. The validated response is scored through a task-specific FOM defined in Supplementary Note 1, which maps the primary physical observable of each device (resonance Q-factor, diffraction efficiency, Strehl ratio, beamlet uniformity, or stopband rejection) to a composite score on a 0–100 scale. Fabrication robustness is assessed by subjecting each design to perturbation testing, broadband and wide-angle operation testing, and other device-specific stability checks. The retention of performance under these tests is scored on a separate 0–100 robustness scale, as detailed in Supplementary Note 2. To demonstrate the advantages of GCCE, we apply the identical FOM and robustness metrics to designs produced by three baseline configurations (single-agent self-aware, multi-agent collaboration and GCCE), with computational effort normalized through the operation-counting methodology. Per-round evolution data across five independent runs, including FOM, robustness, and solver iteration counts for every benchmark, are archived in Supplementary Note 4. To demonstrate the advantages of GCCE, we apply the identical FOM and robustness metrics to designs produced by three baseline configurations (single-agent self-aware, multi-agent collaboration and GCCE). Per-round evolution data across five independent runs, including FOM, robustness, and solver iteration counts for every benchmark, are archived in Supplementary Note 4.

**Operation counting methodology**

To enable quantitative comparison across different agent configurations and design strategies, the framework implements a standardized operation-counting method. Two categories are counted: (1) Orchestrator dispatches. Each occurrence of an agent delegation tag, counted as one operation; (2) specialist agent tool calls. Each invocation of an MCP tool (including simulation, optimization, skill loading, script execution, file I/O, and pipeline steps), counted as one operation. The total operation count is the sum of dispatches and tool calls. Ordinary conversational turns without tool invocation are excluded. This metric provides a hardware-independent measure of computational effort.

**Benchmark tasks**

The generalizability of Photonics-GCCE is evaluated across six representative optical inverse-design tasks:

(1) Ultrahigh-$Q$-factor qBIC metasurface: A $TiO_2$-on-$SiO_2$ bound-state-in-the-continuum resonator requiring simultaneous multi-parameter optimization to maximize the quality factor while maintaining fabricability under documented lithographic tolerances.

(2) Three-band high-pass multilayer filter: A $TiO_2/SiO_2$ thin-film stack satisfying three spectrally separated transmission windows with OD $\geqslant 2$ out-of-band rejection and passband transmission exceeding 90%, requiring coordinated optimization of many layer thicknesses against competing spectral constraints.

(3) One-dimensional anomalous reflector: A Si grating on an Si-$SiO_2$ multilayer atop a glass substrate that routes nearly all incident 1550-nm laser power into the +1 diffraction order while suppressing specular and −1 order leakage.

(4) Polarization-selective two-dimensional beam splitter: A two-dimensional Si structure on an Si-$SiO_2$ multilayer, distributing incident power into prescribed far-field channels.

(5) Beam-focusing metalens: An array of $SiH_x$ nanopillars on $SiO_2$ where continuous phase profiles are discretized onto finite-sized meta-atom elements to achieve diffraction-limited focusing.

(6) 64-level aperiodic DOE beam splitter: A free-form fused-silica element at 532 nm with independently optimized pillar heights across 64 levels, enabling asymmetric far-field intensity patterns.

## Data availability

All data required to evaluate the conclusions of this study are presented in the Article or its Supplementary Information. The dataset is available at https://github.com/Breathay/gcce-data. We have also prepared a supplementary demonstration video, created with the assistance of AI, that provides an animated walkthrough of the Photonics-GCCE design workflow.

**Code availability**

All relevant code is available from the corresponding authors upon reasonable request.

**Acknowledgements**

This work was supported by National Natural Science Foundation of China (62192770, 62575216, 61925504, 62205246, 6201101335, 62020106009, 61621001, 62475192, 62192771, 62192772). National Key R&D Plan of China (2022YFF0604802 and 2023YFF0613600). Science Foundation of Aeronautics (PSSFA) (20240024038001). Shanghai Pilot Program for Basic Research. Science and Technology Commission of Shanghai Municipality (17JC1400800, 20JC1414600, 21JC1406100 and 22ZR1432400). The Special Development Funds for Major Projects of Shanghai Zhangjiang National Independent Innovation Demonstration Zone (ZJ2021-ZD-008). Shanghai Municipal Science and Technology Major Project (2021SHZDZX0100). The Fundamental Research Funds for the Central Universities.

**Author contributions:**

Z.W. and Y.S. initiated the project. W.X. and M.W. constructed the original agents. R.M. conceived the idea of Photonics-GCCE. W.X., R.M., and M.W. developed tools and skills for the Photonics-GCCE, designed the cases and conducted the simulations. D.Z. provided support in physics. J.C. provided support in simulations. Z.W. and S.H. provided DOE beam splitter cases. H.J. provided multilayer filter cases. L.D. and H.L. provided freeform beam splitter cases. Y.L. and H.L. provided anomalous deflection cases. All authors analyzed the data. W.X., R.M., Y.S. and Z.W. prepared the manuscript. A.M., R.F. and C.Q. revised the manuscript. Y.S., Z.W., R.F., C.Q. and X.C. supervised the project.

**Competing interests:**

The authors declare no competing interests.

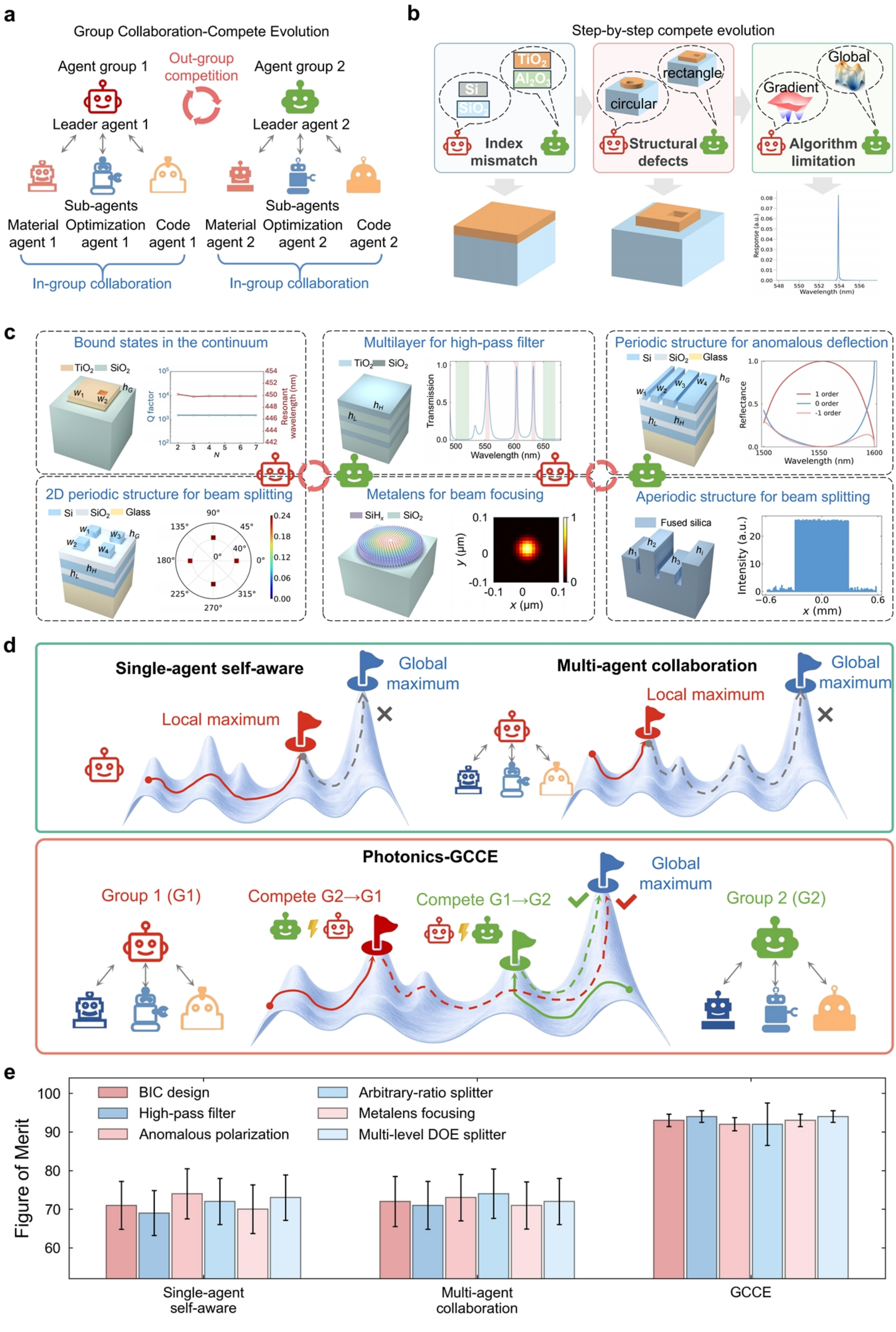
a
Group Collaboration-Compete Evolution
Agent group 1
Out-group competition
Agent group 2
Leader agent 1
Leader agent 2
Sub-agents
Material agent 1
Optimization agent 1
Code agent 1
Material agent 2
Optimization agent 2
Code agent 2
In-group collaboration
b
Step-by-step compete evolution
Index mismatch
Structural defects
Algorithm limitation
circular
rectangle
Gradient
Global
c
Bound states in the continuum
Multilayer for high-pass filter
Periodic structure for anomalous deflection
2D periodic structure for beam splitting
Metalens for beam focusing
Aperiodic structure for beam splitting
Fused silica
d
Single-agent self-aware
Multi-agent collaboration
Local maximum
Global maximum
Photonics-GCCE
Group 1 (G1)
Compete G2→G1
Compete G1→G2
Group 2 (G2)
e
Figure of Merit
BIC design
High-pass filter
Anomalous polarization
Arbitrary-ratio splitter
Metalens focusing
Multi-level DOE splitter
Single-agent self-aware
Multi-agent collaboration
GCCE

**Fig. 1. Photonics-GCCE framework and benchmark portfolio.** **a**, Schematic of group collaboration–compete evolution (GCCE). Two isolated agent groups (each led by a leader agent with material, optimization, and code sub-agents) collaborate within the group while competing across groups. **b**, Step-by-step compete evolution showing how cross-evaluation identifies and corrects recurrent failure modes, including refractive-index mismatch (materials), structural defects (geometry), and algorithmic limitations (optimization), which are then distilled into reusable skills. **c**, Six benchmark tasks used to probe generality: bound states in the continuum (BIC), high-pass multilayer filter, anomalous deflection, beam splitting (2D periodic), metalens focusing, and multi-level diffractive optical element (DOE) splitting. **d**, Search strategy comparison on an optimization landscape. Single-agent and multi-agent configurations (top) are trapped at local maxima. Photonics-GCCE (bottom) combines intra-group collaboration with inter-group competition between adversarial groups G1 and G2, converging to the global maximum (blue flag). **e**, Figure of merit comparison across the three configurations for all six benchmarks, showing monotonic improvement from single-agent through multi-agent collaboration to full GCCE.

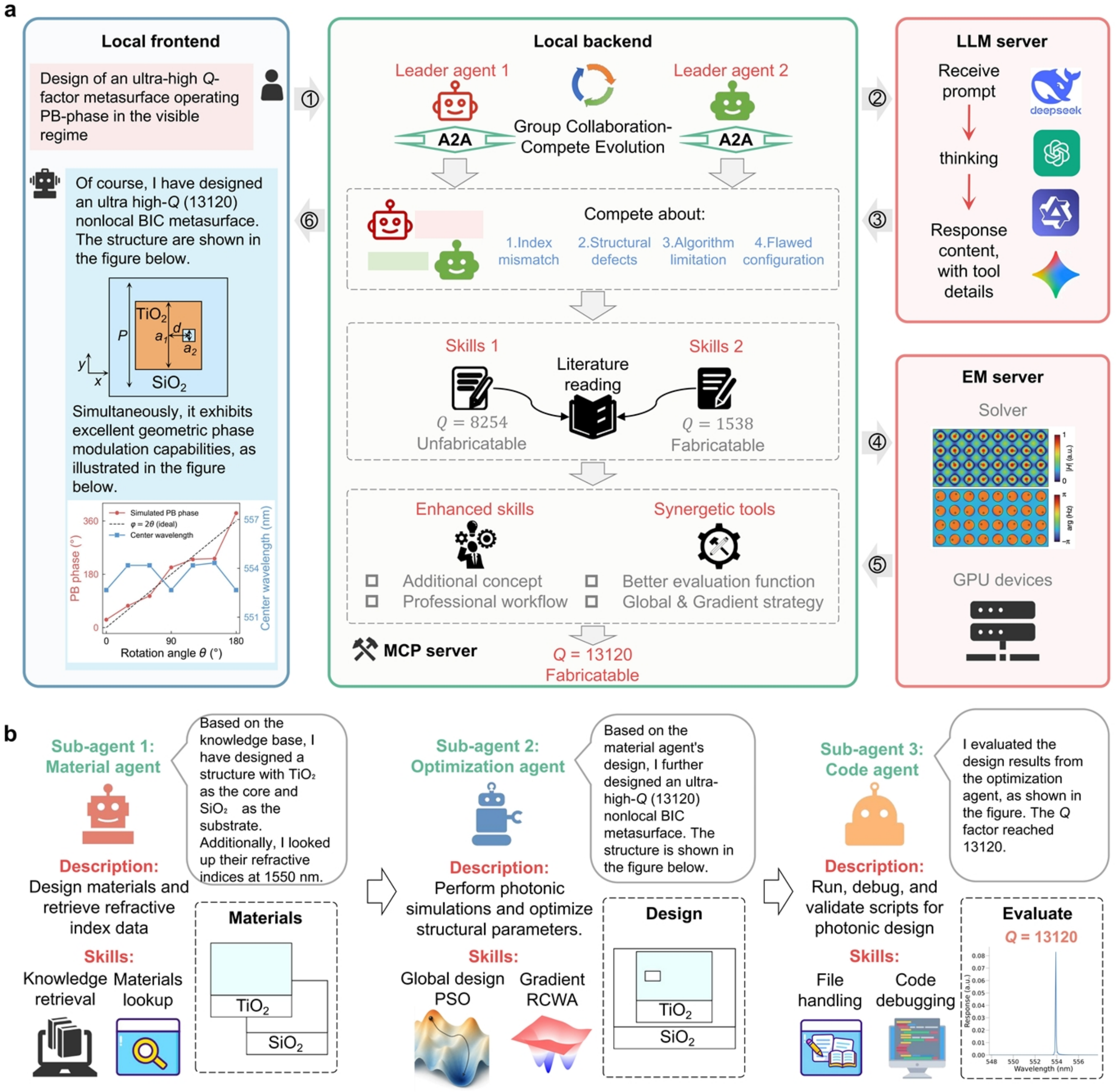

**Fig. 2. System architecture and workflow of Photonics-GCCE. a**, End-to-end deployment of Photonics-GCCE. A local frontend hosts the user interaction; a local backend runs two leader agents communicating via A2A and competing on four axes (index mismatch, structural defects, algorithm limitation, and flawed configuration). LLM servers generate tool-augmented responses, while EM solvers run on GPU devices; experiential skills are accumulated and upgraded through literature reading and post-evaluation. Representative quality factors are shown for each skill set and the final outcome (including fabricability). **b**, Specialist-agent roles within each group. The material agent retrieves optical constants and proposes material stacks; the optimization agent runs photonic simulations and structural optimization; the code agent executes, debugs, validates scripts, and performs independent evaluation to ensure reliable reported metrics.

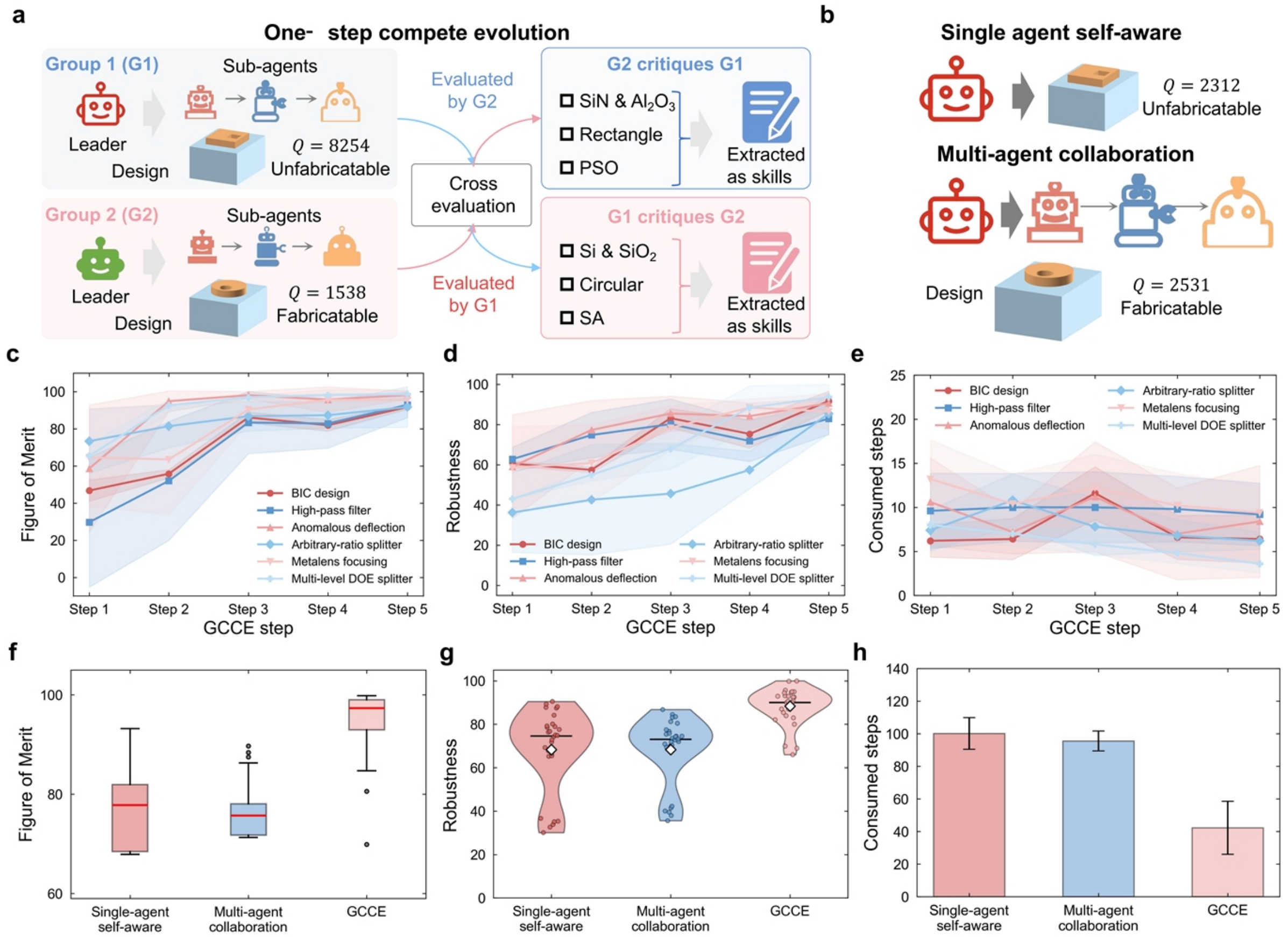

**Fig. 3. Baseline comparisons and evolution dynamics of GCCE. a**, One-step compete evolution. Group 1 proposes a high-$Q$ rectangular design ($Q$ = 8254) that is unfabricatable, while Group 2 proposes a fabricatable circular design ($Q$ = 1538); cross-evaluation extracts actionable skills (materials, geometry, and optimizers such as PSO or SA). **b**, Baseline agent configurations: single self-aware agent, and multi-agent collaboration pipeline. **c–e**, Step-wise evolution across five GCCE steps for six benchmarks, showing monotonic improvement in composite score and robustness alongside reduced consumed steps. **f–h**, Aggregated comparisons across three configurations. GCCE achieves the highest composite score and robustness with the fewest consumed solver iterations, outperforming both the single-agent self-aware and multi-agent collaboration baselines.

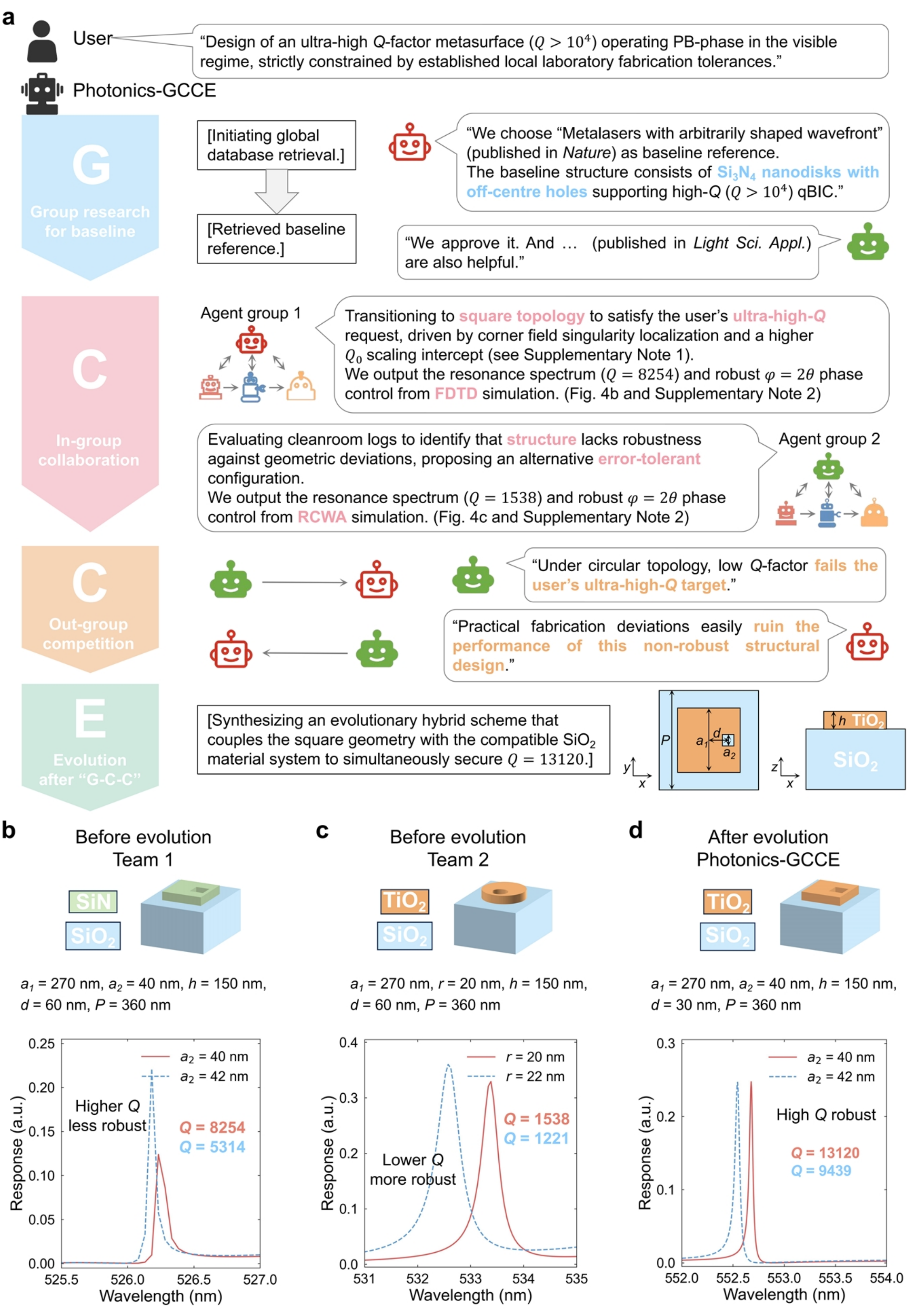

**Fig. 4. Worked qBIC example of Photonics-GCCE. a**, Visible-regime ultra-high-Q brief executed through group research (G), in-group collaboration (C), inter-group competition (C), and evolution (E). **b**, Team 1 ($Si_3N_4/SiO_2$ square nanopillar): target geometry, $Q$ = 8254 (526

nm); 2-nm $a_2$ offset, $Q$ = 5314—high $Q$-factor, low robustness. **c**, Team 2 ($TiO_2/SiO_2$ circular nanodisk with off-center hole): target geometry, $Q$ = 1538 (533 nm); 2-nm inner-radius offset, $Q$ =1221—lower $Q$-factor, higher robustness. **d**, Evolved result (square $TiO_2$ nanopillar with off-center hole on $SiO_2$): target geometry, $Q$ = 13120 (553 nm); same $a_2$ offset, $Q$ = 9439—ultrahigh $Q$-factor with preserved robustness ($a_1$ = 270 nm, $a_2$ = 40 nm, $h$ = 150 nm, $d$ = 30 nm, $P$ = 360 nm).

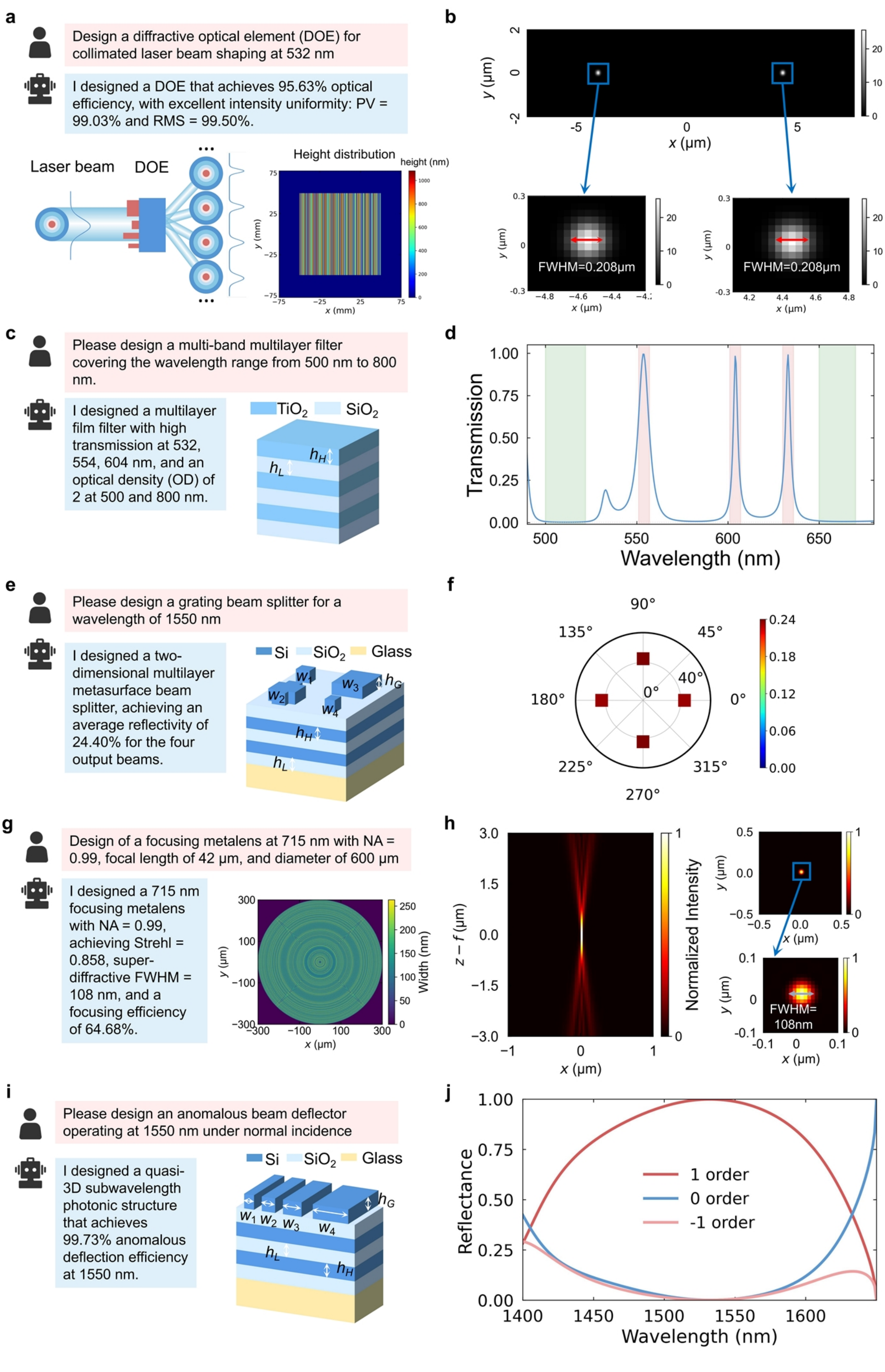
a
Design a diffractive optical element (DOE) for collimated laser beam shaping at 532 nm
I designed a DOE that achieves 95.63% optical efficiency, with excellent intensity uniformity: PV = 99.03% and RMS = 99.50%.
Laser beam
DOE
Height distribution
height (nm)
b
FWHM=0.208μm
FWHM=0.208μm
c
Please design a multi-band multilayer filter covering the wavelength range from 500 nm to 800 nm.
I designed a multilayer film filter with high transmission at 532, 554, 604 nm, and an optical density (OD) of 2 at 500 and 800 nm.
TiO2
SiO2
d
Transmission
Wavelength (nm)
e
Please design a grating beam splitter for a wavelength of 1550 nm
I designed a two-dimensional multilayer metasurface beam splitter, achieving an average reflectivity of 24.40% for the four output beams.
Si
SiO2
Glass
f
g
Design of a focusing metalens at 715 nm with NA = 0.99, focal length of 42 μm, and diameter of 600 μm
I designed a 715 nm focusing metalens with NA = 0.99, achieving Strehl = 0.858, super-diffractive FWHM = 108 nm, and a focusing efficiency of 64.68%.
Width (nm)
h
Normalized Intensity
FWHM=
108nm
i
Please design an anomalous beam deflector operating at 1550 nm under normal incidence
I designed a quasi-3D subwavelength photonic structure that achieves 99.73% anomalous deflection efficiency at 1550 nm.
Si
SiO2
Glass
j
Reflectance
Wavelength (nm)
1 order
0 order
-1 order

**Fig. 5. Additional outcome-only engineering tasks beyond the qBIC case. a-b**, 64-order DOE beam splitter at 532 nm with aperiodic structures of different heights: height-map design and far-field intensity readout, reporting 95.63% optical efficiency with PV = 99.03% and RMS = 99.50% intensity uniformity. **c**-**d**, Multilayer 3-passband high-pass filter: stack schematic and transmission spectrum with five targeted passbands and deep out-of-band rejection. **e-f**, Two-dimensional beam splitter with prescribed intensity allocation at 1550 nm. Structure schematic and far-field pattern, showing a per-channel efficiency of 25.4% across output channels. **g**-**h**, Focusing metalens at 715 nm with NA = 0.99: subwavelength structure and high-NA focusing, achieving Strehl = 0.858, super-diffractive FWHM = 108 nm, and a focusing efficiency of 64.68%. **i**-**j**, One-dimensional ultra-high efficiency perfect anomalous deflector at 1550 nm: subwavelength structure and order-resolved reflectance, achieving 99.73% anomalous deflection efficiency into the +1 order.